\documentclass[12pt]{amsart}
\usepackage{graphicx}

\usepackage{color}

\usepackage{graphicx}
\usepackage{epsfig}

\newtheorem{thm}{Theorem}[section]

\usepackage{amsmath}
\usepackage{amsxtra}
\usepackage{amscd}
\usepackage{amsthm}
\usepackage{amsfonts}
\usepackage{amssymb}
\usepackage{eucal}
\newcommand{\bmb}{\left( \begin{array}{rr}}
\newcommand{\enm}{\end{array}\right)}

\newcommand{\C}{{\mathbb C}}
\newcommand{\Z}{{\mathbb Z}}

\newcommand{\N}{{\mathbb N}}

\newcommand{\bn}{{\mathbf n}}

\newcommand{\bx}{{\mathbf x}}

\newcommand{\bu}{{\mathbf u}}

\newcommand{\al}{{\alpha}}

\numberwithin{equation}{section}

\begin{document}

\title{Integrable Combinatorics}
\author{Philippe Di Francesco}

\address{
Department of Mathematics, University of Illinois at Urbana-Champaign \\
and \\
Institut de Physique Th\'eorique du Commissariat \`a l'Energie Atomique, 
Universit\'e Paris-Saclay\\
e-mail:philippe@illinois.edu
}

\begin{abstract}
We explore various combinatorial problems mostly borrowed from physics, that share the property of
being continuously or discretely integrable, a feature that guarantees the existence of conservation laws
that often make the problems exactly solvable. We illustrate this with: random surfaces, lattice models, and 
structure constants in representation theory. 
\end{abstract}

\maketitle
\date{\today}


%

\section{Introduction}

In this note we deal with combinatorial {\it objects}, mostly provided by physical systems or models. 
These are: random surfaces, lattice models, and structure constants. 
We will illustrate how to solve the various problems, mostly of exact or asymptotic 
enumeration, via a panel of techniques borrowed from pure combinatorics as well as statistical physics. 
The {\it tools} utilized are: generating functions, transfer matrices, bijections, matrix integrals, determinants, field theory, etc. 

We have organized this collection of problems according to some common or analogous properties, essentially
related to their underlying symmetries. Among them the most powerful is the notion of {\it integrability}. The latter appears under many different guises. The first form is continuous: Existence of conservation laws, flat connections, commuting transfer matrices, links to the Yang-Baxter equation, infinite dimensional algebra symmetries. The second form is discrete:
Existence of discrete integrals of motion in discrete time. 

What kind of results did we obtain? Solving a system completely usually entails a complete understanding of correlation functions within the model. This can be achieved by explicit diagonalization of the transfer matrix or Hamiltonians, explicit computation of generating functions, or derivation of complete systems of equations for averaged quantities. As usual in statistical physics, one also investigates the asymptotic (or theormodynamic) properties of the systems, leading to such results as
asymptotic enumeration, identification of phases and their separations, identification of underlying field theoretical descriptions of fluctuations.

One of the main features common to all the problems listed above is some kind of connection to discrete {\it paths} or 
{\it trees}, the two simplest and most fundamental combinatorial objects. The constructs of this note place these two main characters in new non-standard contexts which shed some new light on their deep significance. Together they form the basis of the notion of combinatorial integrability, i.e. the properties shared by combinatorial problems that connect them to discrete or continuous integrable systems.

The paper is organized as follows.
In Section \ref{RSsec}, we explore discretized models of random surfaces, whether  
Lorentzian in 1+1 dimensions (Section \ref{lorgrasec}), or Euclidian in 2 dimensions (Section \ref{2dqgsec}). 
Both type of models display integrability respectively via commuting transfer matrices and discrete integrals of motion, 
which allows to solve them explicitly.

In Section \ref{LMsec}, we first describe the 6 vertex model and its many combinatorial wonders
(Sect. \ref{6vsec}), among which a description of Alternating Sign Matrices (ASM), and links to special types 
of plane partitions, as well as the geometry of nilpotent matrix varieties.

Section \ref{Spinsec} focuses on Lie algebraic structures with a description of Whittaker vectors (Section \ref{whittasec})
using path models, and of graded multiplicities in tensor products occurring in inhomogeneous quantum spin chains with Lie symmetry (Section \ref{fusionsec}). The description of the latter involves a construction of difference operators that generalize
the celebrated Macdonald operators, and can be understood within the context of polynomial representations of Double-Affine Hecke Algebras (DAHA), quantum toroidal algebras, and Elliptic Hall Algebras (EHA).

Finally we gather some important open problems in Section \ref{OPsec}, which we think should shape the future of integrable combinatorics.

\noindent{\bf Acknowledgments.} Much of the work described in this note is the fruit of collaborations with various authors:
R. Behrend, O. Golinelli, E. Guitter, C. Itzykson, R. Kedem, M. Lapa, N. Reshetikhin, R. Soto Garrido, 
P. Zinn-Justin, J.-B. Zuber. 
I also benefited from many important discussions with fellow mathematicians and physicists: O. Babelon, A. Borodin,
J.-E. Bourgine, I. Cherednik, I. Corwin, S. Fomin, M. Gekhtmann, M. Jimbo, R. Kenyon, M. Kontsevich, Y. Matsuo, 
G. Musiker, A. Negut, A. Okounkov, V. Pasquier,  M. Shapiro, O. Schiffman, V. Retakh. 
I acknowledge support by the Morris and Gertrude Fine foundation.

\section{Random surfaces}\label{RSsec}

\subsection{1+1-dimensional Lorentzian triangulations and (continuous) integrability}\label{lorgrasec}
Lorentzian triangulations \cite{LORGRA} are used as a discrete model for quantum gravity in one (space)+1 (time) dimension.
Pure gravity deals with fluctuations of such bare space-times, while matter theories include for instance
particle systems in interction defined on such space-times. 
General relativity expresses the relation between those fluctuations and in particular the associated fluctuations of the
metric, area and curvature of the space-time and the matter stress tensor.
The model for a fluctuating 1+1D space-time is 
an arrangement of triangles organized into time slices as depicted below:
$$ \raisebox{-1.cm}{\hbox{\epsfxsize=7.cm \epsfbox{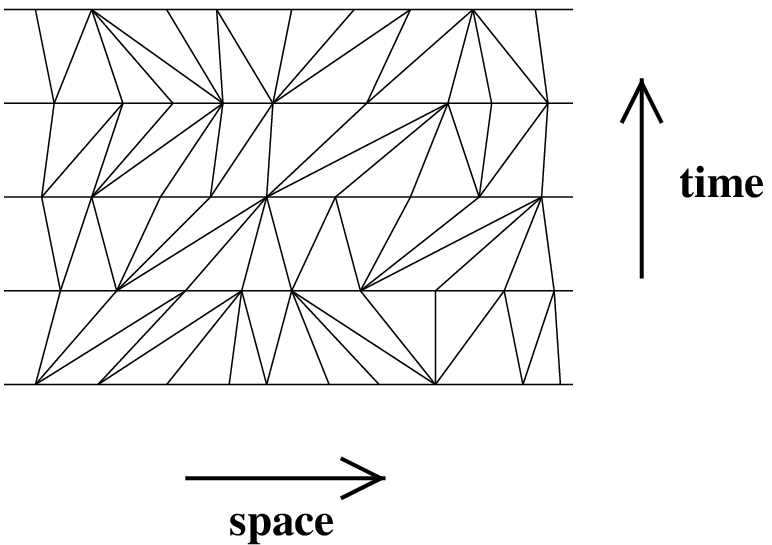}}}  $$
Fluctuations of space are represented by random arrangements of triangles in each time slice, while the time
direction remains regular. These triangulations are best described in the dual picture by considering triangles as vertical half-edges
and pairs of triangles that share a time-like (horizontal) edge as vertical edges between two consecutive time-slices.
We may now concentrate on the transition between two consecutive time-slices which typically looks like:
\begin{equation}\label{transmatbare} \raisebox{-1.cm}{\hbox{\epsfxsize=7.cm \epsfbox{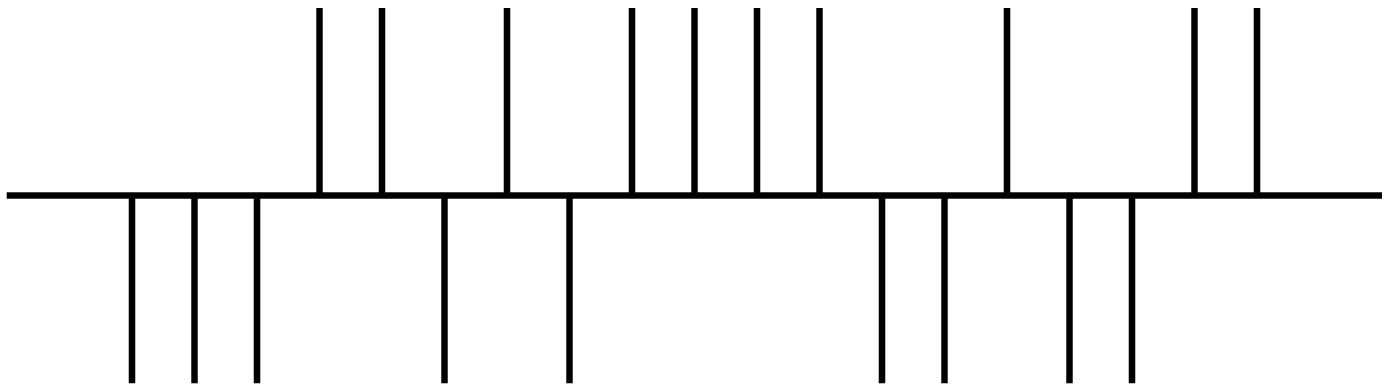}}}  \end{equation}
with say $i$ half-edges on the bottom and $j$ on the top (here for instance we have $i=9$ and $j=10$). 
To take into account the {\it area} and {\it curvature} of space-time, we may introduce a Boltzmann weight $g$ per triangle 
(i.e. per trivalent vertex in the dual picture) and a weight $a$ per pair of consecutive triangles in a time-slice pointing 
in the same direction (both up or both down).
The total weight of a configuration is the product of all local weights pertaining ot it.
It is easy to see that these weights correspond to a transfer operator ${\mathcal T}(g,a)$ which 
describes the configurations of one time-slice with a total $i$ of up-pointing triangles and $j$ of down-pointing ones. The
matrix element between states $i$ and $j$ reads: 
$$T(g,a)_{i,j}= (ag)^{i+j}\, \sum_{k=0}^{{\rm Min}(i,j)} {i\choose k}{j\choose k} a^{-2k} \qquad (i,j\geq 0)$$
Equivalently, the double generating function for matrix elements of ${\mathcal T}(g,a)$ reads:
\begin{equation}\label{simple} f_{T(g,a)}(z,w)=\sum_{i,j\geq 0} T(g,a)_{i,j} z^i w^j =\frac{1}{1-ga(z+w)-g^2(1-a^2)z w} 
\end{equation}
This model turns out to provide one of the simplest examples of quantum integrable system, with an infinite
family of commuting transfer matrices. Indeed, we have:

\begin{thm}[\cite{LORGRA}]
The transfer matrices $T(g,a)$ and $T(g',a')$ commute if and only if the parameters $(g,a,g',a')$ are such that
$\varphi(g,a)=\varphi(g',a')$ where:
$$ \varphi(g,a)=\frac{1-g^2(1-a^2)}{a g} $$
\end{thm}

This and the explicit generating function \eqref{simple} were extensively used in \cite{LORGRA} to diagonalize 
${\mathcal T}(g,a)$ and to compute correlation
functions of boundaries in random Lorentzian triangulations. 

For suitable choices of boundary conditions, the dual random Lorentzian triangulations introduced above may
be viewed as random plane trees. This is easily realized by gluing all the bottom vertices of successive parallel vertical edges
(no interlacing with the neighboring time slices).
A typical such example reads:
$$ \raisebox{-1.cm}{\hbox{\epsfxsize=10.cm \epsfbox{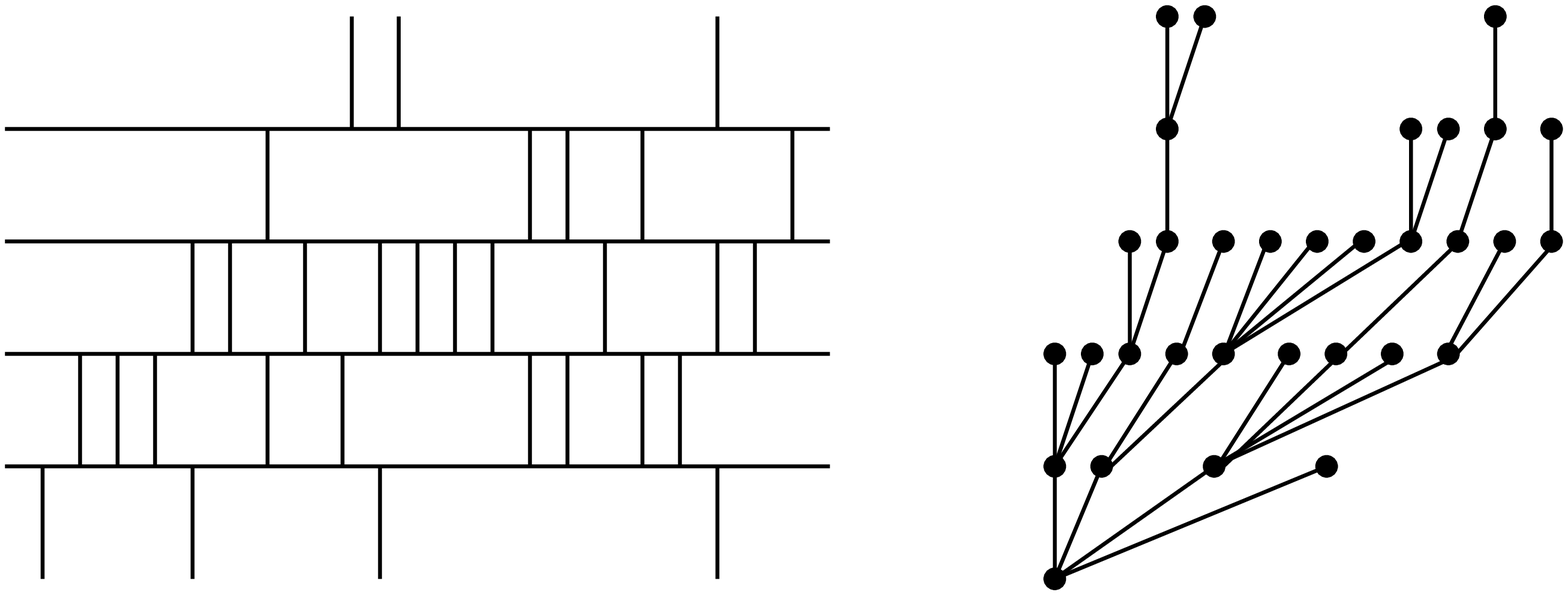}}}$$
Note that the tree is naturally rooted at its bottom vertex.

To summarize, we have unearthed some integrable structure attached naturally to plane trees, one of the most fundamental
objects of combinatorics. Note that in tree language the weights are respectively $g^2$ per edge, and $a$ 
per pair of consecutive descendent edges and per pair of consecutive leaves at each vertex (from left to right).

\subsection{2-dimensional Euclidian tessellations and (discrete) integrability}\label{2dqgsec}

As opposed to Lorentzian gravity, the 2D Euclidian theory involves fluctuations of both space and time, allowing for space-times that
look like random surfaces of arbitrary genus. Those are discretized by tessellations. A powreful tool for enumerating those maps
was provided by matrix integrals, allowing to keep track of the area, as well as the genus via the size $N$ of the matrices 
(see Ref.\cite{DGZ} and references therein).
In a parallel way,
the field-theoretical descriptions of the (critical) continuum limit of two-dimensional quantum gravity (2DQG) have blossomed into a
more complete picture with identification of relevant operators and computation of their correlation functions \cite{corDK}.
This was finally completed by an understanding in terms of the intersection theory of the moduli space of curves with punctures
and fixed genus \cite{KONTS}. Remarkably, in all these approaches a common integrable structure is always present. It takes
the form of commuting flows in parameter space. However, a number of issues were left unadressed by the matrix/field theoretical 
approaches. What about the intrinsic
geometry of the random surfaces? Correlators must be integrated w.r.t. the position of their insertions, leaving us
only with topological invariants of the surfaces. But how to keep track for instance of the geodesic distances between two
marked points of a surface, while at the same time summing over all surface fluctuations?

Answers to these questions came from a better combinatorial understanding of the structure of the (planar) tessellations 
involved in the discrete models. And, surprisingly, yet another form of integrability appeared. Following pioneering
work of Schaeffer \cite{SCH}, it was observed that all models of discrete 2DQG with a matrix model solution 
(at least in genus 0) could be expressed as statistical models of (decorated) trees, and moreover, the decorations
allowed to keep track of geodesic distances between some faces of the tessellations. Marked planar tessellations 
are known as rooted planar maps in combinatorics. They correspond to connected graphs (with vertices, edges, faces)
embedded into the Riemann sphere. Such maps are usually represented on a plane with a distinguished face 
``at infinity", and a marked edge adjacent to that face. The degree of a vertex is the number of distinct half-edges
adjacent to it, the degree of a face is the number of edges forming its boundary.

Consider the example of tetravalent (degree 4) planar maps with 2 univalent (degree 1) vertices, 
one of which is singled out as the root. The Schaeffer bijection associates to each of these a unique rooted tetravalent 
(with inner vertices of degree 4) tree called blossom-tree, with two types
of leaves (black and white), and such that there is exactly one black leaf attached to each inner vertex:
$$  \raisebox{-1.cm}{\hbox{\epsfxsize=10.cm \epsfbox{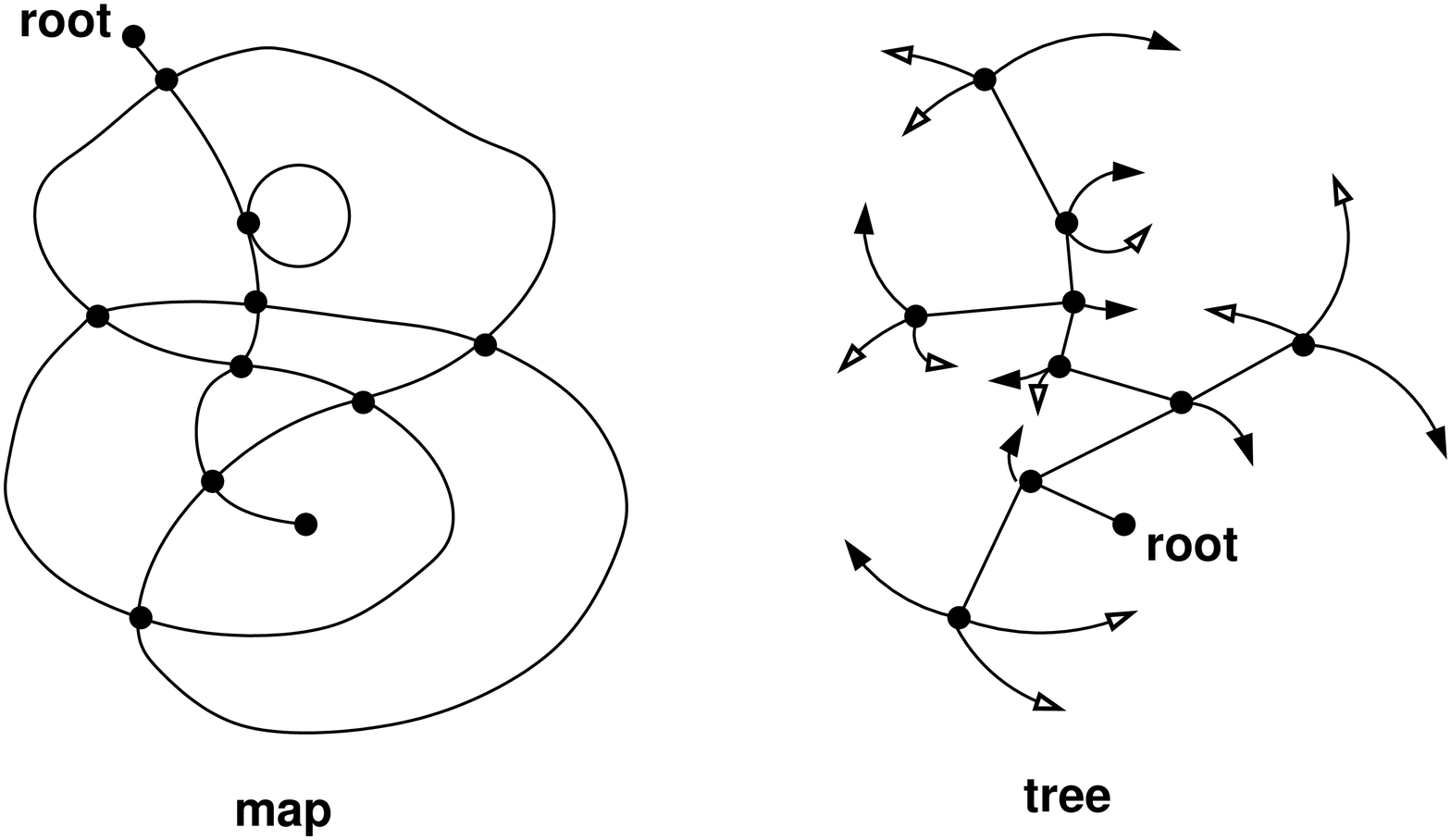}}}$$
This is obtained by the following cutting algorithm: travel clockwise along the bordering edges of the face at infinity, starting from
the root. 
For each traversed edge, cut it if and only if after the cut, the new graph remains connected, 
and replace the two newly formed half-edges
by a black and a white leaf respectively in clockwise order. Once the loop is traveled, this has created a larger face at infinity.
Repeat the procedure until the graph has only one face left: it is the desired blossom-tree, which we reroot at the other
univalent vertex, while the original root is transformed into a white leaf.

This bijection allows to keep track of the geodesic distance between the 2 univalent vertices.
Defining $R_n(g)$ to be the generating function for maps with geodesic distance $\leq n$ 
between the two univalent vertices, we have the following recursion relation \cite{GEOD}:
\begin{equation}\label{crucial}
R_n(g)= 1+ gR_n(g)\left( R_{n+1}(g)+R_n(g)+R_{n-1}(g)\right) 
\end{equation}
easily derived by inspecting the environment of the vertex attached to the root of the tree when it exists. 
It is supplemented by boundary conditions $R_{-1}(g)=0$ and $\lim_{n\to\infty} R_n(g)=R(g)=\frac{1-\sqrt{1-12g}}{6g}$, 
the generating function of maps with no geodesic distance constraint. 

Equation \eqref{crucial}, viewed as governing the evolution of the quantity $R_n(g)$ in the
discrete time variable $n$,  is a classical {\it discrete integrable system}. By this we mean that it has a
{\it discrete integral of motion}, expressed as follows. The function $\phi(x,y)$ defined by
\begin{equation}\label{cons}\phi(x,y)= x y (1-g(x+y))-x-y \end{equation}
is such that for any solution $S_n$ of the recursion relation \eqref{crucial}, the quantity 
$\phi(S_n,S_{n+1})$ is independent of $n$. In other words, the quantity  $\phi(S_n,S_{n+1})$ is 
conserved modulo \eqref{crucial}. (This is easily shown by factoring $\phi(S_n,S_{n+1})-\phi(S_{n-1},S_n)$.).
This conservation law gives in particular a relation of the form: 
$ \phi(R_n(g),R_{n+1}(g))=\lim_{m\to\infty} \phi(R_m(g),R_{m+1}(g))=\phi(R(g),R(g))$. It turns out that we can
solve explicitly for $R_n(g)$:

\begin{thm}[\cite{GEOD}]\label{geodthm}
The generating function $R_n(g)$ for rooted tetravalent planar maps with two univalent vertices
at geodesic distance at most $n$ from each other reads:
$$ R_n(g)=R(g) \frac{(1-x(g)^{n+1})(1-x(g)^{n+4})}{(1-x(g)^{n+2})(1-x(g)^{n+3})} $$
where $x(g)$ is the unique solution of the equation:
$ x+\frac{1}{x}+4=\frac{1}{g R(g)^2} $ 
with a power series expansion of the form $x(g)=g+O(g^2)$.
\end{thm}

The form of the solution in Theorem \ref{geodthm} is that of a discrete soliton with tau-function $\tau_n=1-x(g)^n$.
Imposing more general boundary conditions on the equation \eqref{crucial} leads to elliptic solutions of the same flavor.
The solution above and its generalizations to many classes of planar maps \cite{ramaPDF} have allowed for a better understanding
of the critical behavior of surfaces and their intrinsic geometry. Recent developments include planar three-point
correlations, as well as higher genus results.

To summarize, we have seen yet another integrable structure emerge in relation to (decorated) trees.
This is of a completely different nature from the one discussed in Section \ref{lorgrasec}, 
where a quantum integrable structure
was attached to rooted planar trees. Here we have a discrete classical integrable system, with soliton-like solutions.

\section{Lattice models}\label{LMsec}

\subsection{The six-vertex model and beyond}\label{6vsec}

The Six Vertex (6V) model is the archetypical example of 2D integrable lattice model. It is defined on domains 
of the square lattice $\Z^2$, with configurations obtained by orienting all the nearest neigbor edges 
in such a way that there are exactly to ingoing and two outgoing edges incident to each vertex in the interior of the domain
(ice rule). This gives rise to ${4\choose 2}=6$ local vertex configurations, to which one usually attaches Boltzmann weights.
The integrability of the model becomes manifest if we parametrize these weights with rapidities (spectral parameters) that
are derived from the relevant R-matrix solution of the Yang-Baxter equation. This ensures that the system has an infinite 
set of commuting transfer matrices, similarly to Section \ref{lorgrasec}. This property ensures that the transfer matrix
is explicitly diagonalizable by means of Bethe Ansatz techniques. Note that a certain limit of the transfer matrix yields the
Hamiltonian of the anisotropic XXZ spin chain.

\begin{figure}
\centering
\includegraphics[width=15.cm]{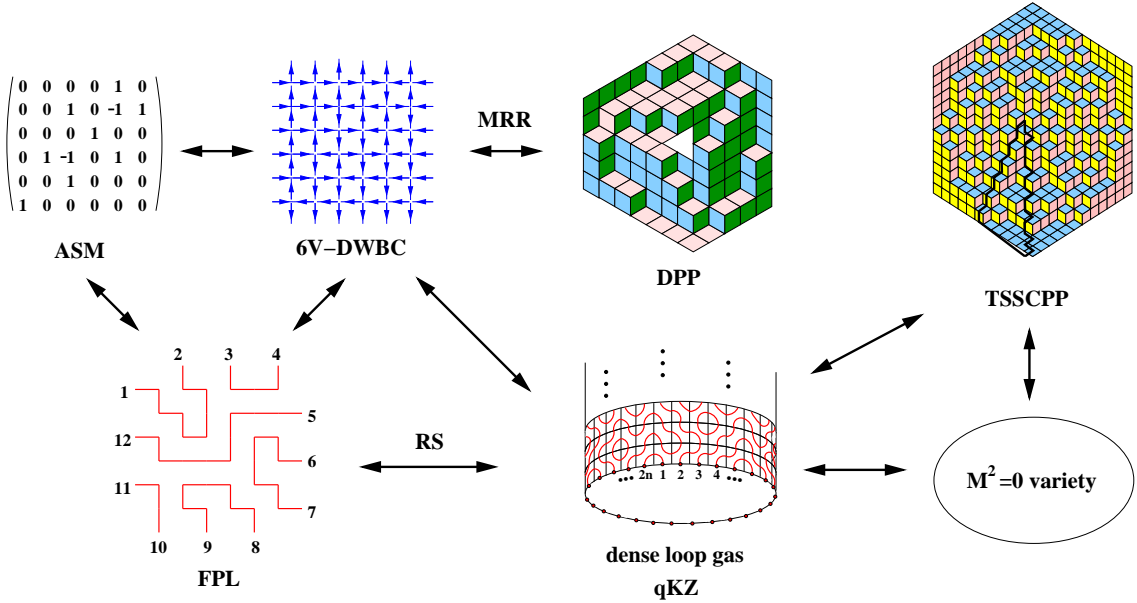}
\caption{\small The combinatorial family of ASMs. From left to right: ASM, 6V-DWBC and FPL , all in bijection;
dense O(n) loop gas: its groundstate/limiting probability vector satisfies the qKZ equation, the components measure 
FPL correlations (RS conjecture); DPP: their refined evaluation matches that of ASMs;
TSSCPP: their refined enumeration matches a sum rule for qKZ solutions at generic $q$ and $z_i=1$; Variety $M^2=0$: 
its degree/multidegree matches solutions of qKZ for $q=1$.}
\label{fig:everything}
\end{figure}

A remarkable web of connections between many combinatorial objects relates to the 6V model, as shown in Fig.\ref{fig:everything}.
The configurations of the 6V model on a square grid of size $n$ with the so-called Domain Wall Boundary 
Conditions (DWBC) that all boundary horizontal arrows are pointed towards the grid while all boundary vertical arrows 
point outward, are in bijection with Alternating Sign Matrices (ASM), namely matrices with entries in $\{-1,0,1\}$ with
alternance of $1$'s and $-1$'s along each row and column, and with row and column sums all equal to $1$. This observation
allowed Kuperberg \cite{KUP} to come up with an elegant proof of the ASM conjecture for the number
of $n\times n$ ASMs, soon after 
the combinatorial proof of Zeilberger \cite{ZEIL}. Another bijection related the configurations of the 6V model with DWBC 
to so-called Fully Packed Loops (FPL) obtained by coloring edges of the square grid in such a way that exactly 2 edges
incident to each inner vertex are colored, while every other boundary edge is colored. The colored edges form closed loops or open
paths connecting boundary edges by pairs (such a pattern of connections is equivalent to non-crossing partitions or link patterns).
The latter remark prompted the celebrated Razumov-Stroganov \cite{RS}
conjecture that FPL configurations with prescribed boundary edge connections
form the Perron-Frobenius eigenvector of the XXZ spin chain at its {\it combinatorial point} (when all Boltzmann weights are $1$),
when expressed in the link pattern basis (in the O(n) model formulation of the spin chain), 
later proved by Cantini and Sportiello \cite{CS}. 
Among the many developments around the conjecture, we used the link between the combinatorial problem and solutions
of the quantum Knizhnik-Zamolodchikov (qKZ) equation for the O(n) model \cite{DZJ0,DZJ1}, which led us to connections 
with the geometry of the  variety of square zero matrices \cite{DZJ2}. Beyond bijections other sets of combinatorial objects 
have the same cardinality $A_n$.
These are the Totally Symmetric Self-Complementary Plane Partitions (TSSCPP) on one hand and the Descending Plane Partitions
(DPP) on the other. Both classes of objects can be formulated as the rhombus tilings of particular domains of the triangular 
lattice with particular symmetries. We found a proof \cite{BDZJ1,BDZJ2} of the Mills-Robbins-Rumsey 
refined ASM-DPP conjecture \cite{MRR}  using generating functions 
similar to \eqref{simple}, however no bijection is known to this day.

It turns out that, among other formulations, the 6V model with DWBC may be expressed as a model of osculating paths, 
namely non-intersecting paths with unit steps $(1,0)$ and $(0,1)$,
from the W border edges to the N border edges of the grid with allowed ``kissing"
or {\it osculating} vertices visited by two paths that do not cross. The latter path formulation allows to predict the arctic curve
phenomenon (i.e. the sharp separation between ordered and disordered phases) for random ASMs \cite{COSPO} 
as well as for random ASMs with a vertical reflection symmetry \cite{DFL}.


\section{Lie algebras, quantum spin chains and CFT}\label{Spinsec}

In this section, we present combinatorial problems/approaches to algebra representation theory. 

\subsection{Whittaker vectors and path models}\label{whittasec}

Whittaker vectors \cite{KOST} are fundamental objects in the representation theory of Lie algebras 
expressed in terms of Chevalley generators $\{e_i,f_i,h_i\}_{i=\epsilon}^r$ and relations
(here $\epsilon=1$ for finite algebras, and $\epsilon=0$ for affine algebras). 
They are instrumental in constructing Whittaker functions, which are eigenfunctions for the
quantum Toda operators, namely Schroedinger operators with kinetic and potential terms coded by the root system of the algebra.
Given a Verma module $V_\lambda= {\mathcal U}(\{f_i\})|\lambda\rangle$ with
highest weight vector $|\lambda\rangle$, a Whittaker vector $v_{\mu,\lambda}$
with parameters $\mu_i$ is an element of the completion of $V_\lambda$ (an infinite series in $V_\lambda$)
that satisfies the relations $e_i v_{\lambda,\mu}=\mu_i v_i$ for all $i$. It is unique upon a choice of normalization.
In Ref.\cite{Whittak} we developed a general approach
to the computation of Whittaker vectors 
by expanding them on the ``words" of the form $f_{i_1}f_{i_2}\cdots f_{i_k}|\lambda \rangle$ 
for arbitrary $i_j\in [\epsilon,r]$ and $k\geq 0$. The latter are of course not linearly independent, but we found some extremely nice and
simple expression for their coefficients $c_{i_1,...,i_k}$ in the expression of $v_{\lambda,\mu}=\sum_{k\geq 0} \sum_{i_1,...,i_k\in [\epsilon,r]} 
c_{i_1,...,i_k}\,f_{i_1}f_{i_2}\cdots f_{i_k}|\lambda \rangle$ (the normalization is chosen so that the empty word has coefficient $c_{0,0,...,0}=1$). 
We made the observation that the set of vectors of the form 
$f_{i_1}f_{i_2}\cdots f_{i_k}|\lambda \rangle$
is in bijection with the set of paths $p\in \mathcal P$ on the positive cone $Q_+$ of the root lattice, 
from the origin to some root $\beta=(\beta_i)_{i\in [\epsilon,r]}$
where $\beta_i$ is the number of occurrences of $f_i$ in the vector (or of the letter $i$ in the word). Indeed, the steps of $p$ 
are taken successively in the directions $i_k,i_{k-1},...,i_1$ in $[\epsilon,r]$. We denote by $|p\rangle =f_{i_1}f_{i_2}\cdots f_{i_k}|\lambda \rangle$.
We have the following general result.

\begin{thm}[\cite{Whittak}]
For finite or affine Lie algebras, the Whittaker vector $v_{\lambda,\mu}$ is expressed as:
$$v_{\lambda,\mu}=\sum_{\beta\in Q_+} \prod_i \mu_i^{\beta_i}  \sum_{{\rm paths}\,p: 0\to \beta} w(p) |p\rangle $$
where the weight $w(p)$ is a product of local weights:
$$w(p)=\prod_{\gamma\in Q_+^*\atop \gamma\ {\rm vertex}\ {\rm of}\ p}\frac{1}{v(\gamma)},\qquad 
v(\gamma)= (\lambda + \rho\vert \gamma)-\frac{1}{2}(\gamma\vert\gamma)$$
\end{thm}

This construction was shown in \cite{Whittak} to extend to the $A$ type quantum algebras ${\mathcal U}_q({\mathfrak sl}_{r+1})$
with local weights depending on both the vertex and the direction of the step from the vertex.

This new formulation of Whittaker vectors yields a very simple proof for the fact that the corresponding 
Whittaker function obeys the quantum Toda equation (classical case), the Lam\'e-like deformed Toda equation (affine, non-critical case)
or the $q$-difference Toda equation (quantum case).

The $q$-Whittaker functions are known to be a degenerate limit of Macdonald polynomials, when $t\to 0$ or $\infty$. This suggests to
look for a possible path formulation of Macdonald polynomials.

\subsection{Fusion product, Q-system cluster algebra and Macdonald theory}\label{fusionsec}

\subsubsection{Graded characters and quantum Q-system}

We now turn to the combinatorial problem of finding the fusion coefficients 
${\rm Mult}_q(\otimes KR_{\al,n}^{\otimes n_{\al,n}};V_\lambda)$
for graded tensor product decompositions of so-called
Kirillov-Reshetikhin \cite{KR} (KR) modules $KR_{\al,n}$ ($\al\in [1,r];n\in \N$) of a Lie algebra into irreducibles. 
The grading, inherited from the loop algebra \cite{FL} (fusion product) turns out to have many
equivalent formulations: as energy of the corresponding crystal, as linearized energy in the Bethe Ansatz solutions
of the corresponding inhomogeneous isotropic XXX quantum spin chain (the physical system at the origin of the problem, from
which so-called fermionic formulas for graded multiplicities were derived \cite{HKOTY}).
Recently, we have found yet another interpretation of this grading, as being provided by the canonical quantum deformation
of the cluster algebra of the so-called Q-system for the algebra \cite{krDFK}.

The latter is a recursive system for scalar variables $Q_{\al,n}$ $\al=1,2...,r$, $n\in \Z$. For the case of $A_r$ it takes the form:
$$
Q_{\al,n+1}\, Q_{\al,n+1}=(Q_{\al,n})^2-Q_{\al+1,n}\, Q_{\al-1,n} 
$$
with boundary conditions $Q_{0,n}=Q_{r+1,n}=1$ for all $n\in \Z$. It is satisfied by the KR characters 
$Q_{\al,n}=\chi_{KR_{\al,n}}(\bx)$.
This is a discrete integrable system: there exist $r$ algebraically independent polynomial quantities of the $Q$'s that are
conserved modulo the system, which we can view as describing evolution of the variables $Q_{\al,n}$ in discrete time 
$n$ \cite{heapDFK,diffDFK}. Taking advantage of this property, we were able to solve such systems by means of 
(strongly) non-intersecting lattice paths (the solution involves also a new continued fraction rearrangement theory
\cite{heapDFK,cfPDF}).

Such systems exist for all finite and affine algebras, and were shown to be particular sets of mutations in some cluster algebras
\cite{clusK,clusDFK}. 
As such, they admit a natural quantization into a $q$-deformed, non-commutative Q-system, coined the quantum Q-system.
For the case $A_r$ it reads:
\begin{equation}
Q_{\al,n}\,Q_{\beta,n+1}=q^{\lambda_{\al,\beta}}\, Q_{\beta,n+1}\, Q_{\al,n}, \quad
q^{\lambda_{\al,\al}}\,Q_{\al,n+1}\, Q_{\al,n+1}=(Q_{\al,n})^2-q\, Q_{\al+1,n}\, Q_{\al-1,n} 
\end{equation}
where $\lambda_{\al,\beta}=(C^{-1})_{\al,\beta}$, $C$ the Cartan matrix of the algebra, and with the boundary conditions
$Q_{0,n}=Q_{r+1,0}=1$, $Q_{r+2,n}=0$ for all $n\in \Z$. 
The non-commuting variables $Q_{\al,n}$ play the role of quantized KR characters. The path solutions of the classical Q-system 
admit a non-commutative version using non-commutative continued fractions \cite{ncDFK}.

For simplicity let us  perform a change of variables. Define
$A=Q_{r+1,1}$ and the degree operator $\Delta$ such that $\Delta \, Q_{\al,n}= q^{\al n} Q_{\al,n}\, \Delta$.
Then the new variables
$M_{\al,n}:= q^{-\frac{1}{2}\lambda_{\al,\al}(n+r+1)}\,Q_{\al,n} \Delta^{\frac{\al}{r+1}}$ are subject to the new ``M-system":
\begin{equation}
M_{\al,n}\,M_{\beta,n+1}=q^{{\rm Min}(\al,\beta)}\, M_{\beta,n+1}\, M_{\al,n},\quad
q^{\al}\,M_{\al,n+1}\, M_{\al,n+1}=(M_{\al,n})^2-M_{\al+1,n}\, M_{\al-1,n}
\end{equation}
with boundary conditions $M_{0,n}=1$ and $M_{r+1,n}=A^n \Delta$.

\begin{thm}
We have the following representation of the M-system via difference operators acting on the ring of symmetric functions of $N=r+1$
variables $(x_1,...,x_N)$:
\begin{equation}\label{difop}
M_{\al,n}=\sum_{I\subset[1,N]\atop |I|=\al} x_I^n \prod_{i\in I\atop j\not\in I} \frac{x_i}{x_i-x_j} \Gamma_I
\end{equation}
where $x_I=\prod_{i\in I}x_i$, $\Gamma_I=\prod_{i\in I}\Gamma_i$, and $\Gamma_i$ is the multiplicative q-shift operator on the $i$-th variable:
$ (\Gamma_i\, f)(x_1,x_2,...,x_N)= f(x_1,...,x_{i-1},q x_i,x_{i+1},...,x_N)$, and 
with moreover
$ A=x_1x_2\cdots x_N$, and $\Delta=\Gamma_1\Gamma_2\cdots \Gamma_N$.
\end{thm}

Let 
$\chi_\bn(q;\bx)$, $\bn=\{n_{\al,n}\}_{\al\in [1,r];n\in [1,k]}$, $\bx=(x_1,...,x_N)$,
denote the graded character i.e. the generating function for graded multiplicities:
$\chi_\bn(q;\bx)=\sum_{\lambda} {\rm Mult}_q(\otimes KR_{\al,n}^{\otimes n_{\al,n}};V_\lambda) \chi_\lambda(\bx)$,
where the irreducible characters $\chi_\lambda(\bx)=s_\lambda(\bx)$ are the Schur functions.

\begin{thm}[\cite{diffDFK}]
The graded character for the tensor product $\otimes KR_{\al,n}^{\otimes n_{\al,n}}$ reads
$$ \chi_\bn(q^{-1};\bx)=q^{-a(\bn)}\, \prod_{j=k}^1 \prod_{\al=1}^r (M_{\al,j})^{n_{\al,j}} \cdot 1$$
with
$ a(\bn)=\frac{1}{2}\sum_{i,j,\al,\beta} n_{\al,i}{\rm Min}(i,j){\rm Min}(\al,\beta)n_{\beta,j} -\frac{1}{2}\sum_{i,\al}i \al \,n_{\al,i}$.
\end{thm}

The results above were so far only derived for the A case, but they can be extended to $B,C,D$ types \cite{WIP}.

\subsubsection{From Cluster algebra to quantum toroidal and Elliptic Hall algebras}

The form of the difference operator \eqref{difop} is reminiscent of the celebrated Macdonald operators \cite{macbook}, 
for which the Macdonald polynomials form a complete family of eigenvectors. These were best understood in the context
of Double Affine Hecke Algebra \cite{Cheredbook}, in the functional representation. We actually found that a certain action 
of the natural $SL_2(\Z)$ symmetry of DAHA produces the following {\it generalized Macdonald} difference operators 
in the functional representation:
\begin{equation}\label{qtdiffop}
{\mathcal M}_{\al,n}=\sum_{I\subset[1,N]\atop |I|=\al} x_I^n \prod_{i\in I\atop j\not\in I} \frac{t x_i-x_j}{x_i-x_j} \Gamma_I
\end{equation}
We note the relation $M_{\al,n}=\lim_{t\to \infty} t^{-\al(N-\al)} \, {\mathcal M}_{\al,n}$. We may therefore identify the quantum cluster
algebra solution of the Q-system with the $t\to \infty$ (so-called dual q-Whittaker) limit of generalized Macdonald operators.

These operators allow to construct a
representation of the so-called Ding-Iohara-Miki \cite{DINGIO,MIKI} (DIM) or quantum toroidal ${\mathfrak gl}_1$ algebra as follows \cite{qtorDFK}.
Introduce the currents:
\begin{equation}
{\mathfrak e}(z):= \frac{q^{1/2}}{1-q}\sum_{n\in \Z} q^{n/2} z^n {\mathcal M}_{1,n} \quad {\rm and}\quad 
{\mathfrak f}(z):={\mathfrak e}(z)\big\vert_{q\to q^{-1},t\to t^{-1}}
\end{equation}
as well as the series
\begin{equation}\label{defpsipm}
\psi^{\pm}(z):=\prod_{i=1}^{N} 
\frac{(1-q^{-\frac{1}{2}}t (z x_i)^{\pm 1})(1-q^{\frac{1}{2}}t^{-1} (z x_i)^{\pm 1})}{(1-q^{-\frac{1}{2}} (z x_i)^{\pm 1})(1-q^{\frac{1}{2}} (z x_i)^{\pm 1})} \in \C[[z^{\pm 1}]]
\end{equation}
\begin{thm}[\cite{qtorDFK}]
The currents and series $({\mathfrak e},{\mathfrak f},\psi^{\pm})$ form a level $(0,0)$ representation of the DIM algebra.
\end{thm}
In particular, we have the following exchange relation:
$$ g(z,w) \, {\mathfrak e}(z)\, {\mathfrak e}(w)+g(w,z) \,{\mathfrak e}(w)\, {\mathfrak e}(z)=0,\qquad g(z,w)=(z-q w)(z-t^{-1}w)(z-q^{-1}t w)$$
In the $t\to \infty$ limit this reduces to the ${\mathcal U}_{\sqrt{q}}(\widehat{{\mathfrak sl}_2})$ upper Borel subalgebra relation, while the 
DIM relations go to some interesting degeneration, directly connected to the quantum Q-system cluster algebra \cite{diffDFK,qtorDFK}.
Algebra relations allow in particular to derive a quantum determinant formula for $M_{\al,n}$ as a polynomial of the $M_n:=M_{1,n}$'s.
Let us introduce the currents ${m}_\al(z):=\sum_{n\in \Z} z^n M_{\al,n}$, 
and in particular $m(z):=m_1(z)=\lim_{t\to \infty} t^{1-N} \frac{1-q}{q^{1/2}} {\mathfrak e}(z)$, then:
\begin{thm}[\cite{qtDFK}]
\begin{equation}\label{qdet}
m_\al(z)= \left\{ \left(\prod_{1\leq i<j\leq \al} \Big(1-q \frac{u_j}{u_i}\Big) \right)
m(u_1)m(u_2)\cdots m(u_\al) \right\} \Bigg\vert_{(u_1u_2\cdots u_\al)^n}
\end{equation}
where the subscript stands for the coefficient of $(u_1u_2\cdots u_\al)^n$.
\end{thm}
Note also that the function of $\bu$ in \eqref{qdet} is a multi-current generating series. Let us define $M_{a_1,...,a_\al}$ to be the coefficient
of $u_1^{a_1}\cdots u_\al^{a_\al}$ in this series.
There is a very nice expression for $M_{a_1,...,a_\al}$ as a quantum determinant, involving a sum 
over Alternating Sign Matrices. 
This is because the quantity $\prod_{i<j}v_i+\lambda v_j$ is the $\lambda$-determinant $\lambda\!\det( V_n)$
(as defined by Robbins and Rumsey \cite{RR})
of the Vandermonde matrix $V_n:=(v_i^{n-j})_{1\leq i,j\leq n}$.
We denote by $ASM_n$ the set of ASM of size $n$.
The inversion number of an ASM is the quantity $I(A)=\sum_{i>k,j<\ell} A_{i,j}A_{k,\ell}$. We also denote by $N(A)$
the number of $-1$'s in $A$. Let us also define the column vector $v=(n-1,n-2,..,1,0)^t$,
and for each ASM $A$ we denote by $m_i(A):= (Av)_i$.
Then we have the explicit formula, obtained by taking $\lambda =-q$ for the $\lambda$-determinant of the 
$\al\times \al$ Vandermonde matrix $V_\al$:
$$\prod_{1\leq i<j\leq \al}(v_i-q v_j)=\sum_{A\in ASM_n} (-q)^{I(A)-N(A)} (1-q)^{N(A)} \prod_{i=1}^n v_i^{m_i(A)} $$
Combining this with \eqref{qdet}, we deduce the following compact expression for the quantum determinant:

\begin{thm}\label{qdethm}
The quantum determinant of the matrix $\left( M_{a_j+i-j}\right)_{1\leq i,j\leq \al}$ reads:
\begin{equation}\label{ASMqdet}
M_{a_1,...,a_\al}=\left\vert \left( M_{a_j+i-j}\right)_{1\leq i,j\leq \al} \right\vert_q=\sum_{A\in ASM_\al}
(-q)^{I(A)-N(A)}(1-q)^{N(A)} \, \prod_{i=1}^\al M_{ a_i+\al-i-m_i(A)}
\end{equation}
\end{thm}

Finally, we use a known isomorphism \cite{SHIVAS} between the Spherical DAHA with the Elliptic Hall algebra (EHA) to make the connection between
generalized Macdonald operators and a functional representation of the EHA \cite{qtDFK}.  From this connection, we obtain new algebraic
relations between the operators ${\mathcal M}_{\al,n}$, inherited from the ``skinny triangle" relations of \cite{BS}.

To conclude, the results of this section are so far valid for the A type only. It would be interesting to investigate the (t-deformed)
algebraic structures hiding behind the B,C,D cases as well.

\section{Open problems}\label{OPsec}

In this note, we described various instances of discrete or continuous integrability in combinatorial problems. A recurrent theme is 
the ability to rephrase said combinatorial problems in terms of paths or trees. 

Paths are very important objects.  
Equipped with non-commutative weights, paths allow to describe non-commutative monomials in finitely generated 
non-commutative algebras. We have encountered a few instances of this in the present note. A crucial question remains open:
how to deal with {\it families} of non-intersecting non-commutative paths? We have found specific answers in the cases where
the non-commutativity is ``under control", e.g. in the case of quantum path weights with specific q-commuting relations
\cite{aonePDF}. More general non-commuting weighted paths can be described via the theory of quasideterminants \cite{GR},
however it remains to find a good theory for non-intersecting non-commutative paths, and perhaps a non-commutative version
of the Lindstr\"om-Gessel-Viennot (LGV) determinant formula.

Interacting paths are a combinatorial version of interacting fermions. Starting from NILP, we can turn on some interaction, by 
for instance allowing paths to touch without crossing (osculating paths) and attaching a contact energy to such instances. 
As shown in the case of the 6V/ASM model, such interactions still allow for solving, together with the machinery of integrable 
lattice models. As another indication, he so-called tangent method for determining phase transitions in large random tilings 
(arctic curves) seems to apply to interacting paths as well. The determinant form of the partition function for the 
6V model with DWBC would point to the fact that there should exist determinant formulas for interacting paths that generalize LGV.
It seems that a number of interacting path problems are still open, and a systematic study is in order.

\end{document}